\documentclass[aps,preprint]{revtex4}%
\usepackage{amsfonts}
\usepackage{amsmath}
\usepackage{amssymb}
\usepackage{graphicx}%
\providecommand{\U}[1]{\protect\rule{.1in}{.1in}}

\begin{document}
\preprint{ }
\title[ ]{Pseudoharmonic oscillator in quantum mechanics\\with a minimal length}
\author{Djamil Bouaziz}
\email{djamilbouaziz@univ-jijel.dz}
\affiliation{Laboratoire de Physique Th\'{e}orique, D\'{e}partement de Physique,
Facult\'{e} des Sciences Exactes et Informatique, Universit\'{e} de Jijel, BP
98, Ouled Aissa, 18000 Jijel, Algeria}
\author{Abdelmalek Boukhellout }
\email{abdelmalek-boukhellout@univ-jijel.dz}
\affiliation{Laboratoire de Physique de Rayonnement et Applications, D\'{e}partement de
Physique, Facult\'{e} des Sciences Exactes et Informatique, Universit\'{e} de
Jijel, BP 98, Ouled Aissa, 18000 Jijel, Algeria}

\begin{abstract}
The pseudoharmonic oscillator potential is studied in non relativistic quantum
mechanics with a generalized uncertainty principle characterized by the
existence of a minimal length scale, $(\Delta x)_{\min}=$ $\hbar\sqrt{5\beta}%
$. By using a perturbative approach, we derive an analytical expression of the
energy spectrum in the first order of the minimal length parameter $\beta$. We
investigate the effect of this fundamental length on the vibration-rotation
energy levels of diatomic molecules through this potential function
interaction. We explicitly show that the minimal length would have some
physical importance in studying the spectra of diatomic molecules.

\end{abstract}
\volumeyear{2013}
\volumenumber{number}
\issuenumber{number}
\eid{identifier}
\date[Date text]{date}
\received[Received text]{date}

\revised[Revised text]{date}

\accepted[Accepted text]{date}

\published[Published text]{date}

\startpage{1}
\maketitle

\section{Introduction}

\bigskip\ Several studies in quatum gravity propose to modify the standard
Heisenberg uncertainty principle to the form: $\left(  \Delta X\right)
\left(  \Delta P\right)  \geq\frac{\hbar}{2}(1+\beta\left(  \Delta P\right)
^{2}+...)$, where $\beta$ is a small positive parameter
\cite{garay,amati,magiore}. This generalized uncertainty principle (GUP)
includes a nonzero minimal uncertainty in position (minimal length), given by
$\left(  \Delta X\right)  _{\min}=\hbar\sqrt{\beta}$. This minimal length is
assumed to be on the order of the Planck length of $10^{-35}$ m, which is a
lower bound to all physical length scales \cite{pad,jaekel,x}. The
introduction of the GUP in quantum theory leads to fundamental consequences on
the mathematical basis of quantum mechanics. One of the most important
implications of this GUP is the deformation of the commutation relation
between position and momentum operators to the form: $[\widehat{X},\widehat
{P}]=i\hbar(1+\beta\widehat{P}^{2})$ \cite{k1}. The formalism, based on this
deformed algebra, together with the new concepts it implies, has been
discussed extensively in one and more dimensions by Kempf and co-workers in
Refs. \cite{k1,k2,k3,k4,k5}.

In recent years, various topics were studied in this modified version of
quantum mechanics, amongst other, we cite, the Schr\"{o}dinger equation for,
the harmonic oscillator potential \cite{k1,brau,chang}, the hydrogen atom
problem \cite{brau,sandor,stetsko,st,boua3}, the singular inverse square
potential \cite{boua1,boua2}, the problem of a particle in a gravitational
quantum well \cite{brau2,nozari}. In relativistic equations, the minimal
length was introduced in the Dirac equation, with a constant magnetic field
\cite{M,dirac}, with vector and scalar linear potentials \cite{cha}, for the
hydrogen atom potential \cite{rh}, and the Dirac oscillator \cite{q,do}. The
effect of the GUP was also investigated in the context of the Klein-Gordon
equation \cite{KG1,KG2,M}. Finally, the Casimir effect \cite{casimir,c}, and
the black body radiation \cite{black-body} have been studied within this
modified formalism of quantum mechanics. For a large number of references in
connection with this subject, see, Ref. \cite{hossenfelder}. For the sake of
completness, let us mention that there is also another form of the GUP,
recently proposed in the literature, which incorporates a minimal length and a
maximal momentum \cite{ali,ali2}. Much attention is equally given to the study
of the implications of such GUP on physical problems, see, for instance, Refs.
\cite{pou,pou2,ching,ching2}.

Only a few problems have been solved exactly in the formalism of quantum
mechanics with a minimal length, such as the Schr\"{o}dinger equation for the
harmonic oscillator \cite{chang} and for the singular inverse square potential
\cite{boua1,boua2}. In the hydrogen atom problem, for instance, the effect of
the minimal length is assumed to be too small and studied perturbatively in
coordinate space \cite{brau,sandor,stetsko,st}.

It is important to note that any theory based on the GUP is supposed to
account for quantum gravitational effects, which arise at scales of the order
of the Planck length \cite{small}. However, it has been argued that, in
relativistic or nonrelativistic quantum mechanics, this elementary length
incorporated in the GUP may be viewed as an intrinsic scale characterizing the
system under study \cite{k2,boua1}. Consequently, the formalism based on these
deformed commutation relations may provide a new model for an effective
description of complex systems such as quasiparticles, nuclei, and molecules
\cite{k2}.

The purpose of this paper is to study another interaction, in quantum
mechanics with a GUP, namely, the molecular pseudoharmonic oscillator (PHO),
$V(r)=D_{e}(\frac{r}{r_{e}}-\frac{r_{e}}{r})^{2}$, where $D_{e}$ is the
dissociation energy and $r_{e}$ is the equilibrium internuclear distance
\cite{R9}. The PHO is one of the most important molecular potentials; it is
especially relevant in studying diatomic molecules, and it is of great
importance in chemical physics, molecular physics and other fields of physics
\cite{R2,R3,p1,p3,p4,p5}. The main feature of the Schr\"{o}dinger equation
with this potential function is that it allows its exact solution for
arbitrary rotational quantum number $\ell$. This is an advantage compared to
other molecular potentials such as the well-known Morse potential \cite{flug},
which can solved exactly only for non-rotating diatomic molecules.

The objective of this work is twofold, first to compute perturbatively the
energy spectrum of the PHO potential in the context of the deformed
Schr\"{o}dinger equation with a minimal length, and second, to investigate the
effect of the deformation parameter on the vibrational and rotational energy
levels of diatomic molecules. Of course, in such application, quantum
gravitational effects are extremely small, and the the minimal length would
not be on the order of the Planck scale but rather it might be associated to a
dimension of the molecule.

The rest of this paper is organized as follows. In Sec. II, we present the
basic equations of quantum mechanics with a GUP. Sec. III is devoted to the
study of the PHO in quantum mechanics with a GUP, and the effect of the
minimal length included in the formalism, on the\ vibration-rotation energy
levels of diatomic molecules. In the last section, we summarize our results
and conclusions.\newpage

\section{\bigskip BASIC EQUATIONS OF QUANTUM MECHANICS WITH A GUP}

We study the PHO potential in deformed quantum mechanics based on the
following 3-dimensional modified Heisenberg algebra \cite{k1,k2}:%

\begin{align}
\left[  \widehat{X}_{i},\widehat{P}_{j}\right]   &  =i\hbar\left[  \left(
1+\beta\widehat{P}^{2}\right)  \delta_{ij}+\beta^{\prime}\widehat{P}%
_{i}\widehat{P}_{j}\right]  ,\nonumber\\
\left[  \widehat{P}_{i},\widehat{P}_{j}\right]   &  =0,\label{4}\\
\left[  \widehat{X}_{i},\widehat{X}_{j}\right]   &  =i\hbar\frac{2\beta
-\beta^{\prime}+\beta\left(  2\beta+\beta^{\prime}\right)  \widehat{P}^{2}%
}{1+\beta\widehat{P}^{2}}\left(  \widehat{P}_{i}\widehat{X}_{j}-\widehat
{X}_{i}\widehat{P}_{j}\right)  .\nonumber
\end{align}

These commutators imply the following generalized uncertainty principle
(GUP):
\begin{equation}
\left(  \Delta X_{i}\right)  \left(  \Delta P_{i}\right)  \geq\frac{\hbar}%
{2}\left(  1+\beta\sum\limits_{j=1}^{3}[\left(  \Delta P_{j}\right)
^{2}+\left\langle \widehat{P}_{j}\right\rangle ^{2}]+\beta^{\prime}[\left(
\Delta P_{i}\right)  ^{2}+\left\langle \widehat{P}_{i}\right\rangle
^{2}]\right)  ,
\end{equation}
which leads to a lower bound of $\Delta X_{i}$, the so-called minimal length,
given by \cite{k2,brau}
\begin{equation}
\left(  \Delta X_{i}\right)  _{\min}=\hbar\sqrt{3\beta+\beta^{\prime}},\text{
\ \ }\forall i. \label{3}%
\end{equation}
A fundamental consequence of the generalized uncertainty priciple is the loss
of localization in coordinate space due to the presence of a nonzero minimal
uncertainty in position measurements, so that, momentum space would be more
convenient for solving any eigenvalue problem.

The most used momentum representation, satisfying\ Eqs. (\ref{4}), is that
given for the first time in Ref. \cite{k2}
\begin{equation}
\widehat{X}_{i}=i\hbar\left(  (1+\beta p^{2})\dfrac{\partial}{\partial p_{i}%
}+\beta^{^{\prime}}p_{i}p_{j}\dfrac{\partial}{\partial p_{j}}+\gamma
p_{i}\right)  ,\text{ }\widehat{P}_{i}=p_{i}, \label{m}%
\end{equation}
where $\gamma$ is related to $\beta$ and $\beta^{\prime}.$

The problem with this representation is that the solution to the deformed
Schr\"{o}dinger equation is not often simple; only a few problems have been
solved exactly in momentum space with a minimal length. For instance, the
hydrogen atom potential, with representation (\ref{m}), can be solved only for
the $s$-waves, in the case $\beta^{\prime}=2\beta$ and in the first order of
$\beta$ \cite{boua3}.

The other important representation used in the literature is the following
coordinate space representation \cite{brau}:%

\begin{equation}
\widehat{X}_{i}=\widehat{x}_{i},\text{ \ \ \ }\widehat{P}_{i}=\widehat{p}%
_{i}\left(  1+\beta\widehat{p}^{2}\right)  , \label{brau}%
\end{equation}
which is valid in the particular case $\beta^{\prime}=2\beta$, up to the first
order of $\beta$. The operators $\widehat{x}_{i}$ and $\widehat{p}_{i}$
satisfy the standard commutation relations of ordinary quantum mechanics.

The main advantage of representation (\ref{brau}) is that it allows to apply
the perturbation theory in order to study the Schr\"{o}dinger equation with a
given interaction.

In the following section we will study the Shr\"{o}dinger equation for the PHO
molecular potential using representation (\ref{brau}), we will be interested
in particular to the effect of the minimal length on the energy spectrum.

\section{PHO Potential in quantum mechanics with a GUP}

While several potentials were studied in quantum mechanics with a GUP, here we
consider the PHO molecular potential, which has the form
\begin{equation}
V(r)=D_{e}(\frac{r}{r_{e}}-\frac{r_{e}}{r})^{2}, \label{p}%
\end{equation}
where $D_{e}$ is the dissociation energy and $r_{e}$ is the equilibrium
internuclear distance of a given diatomic molecule \cite{R9}.

As is clear, the potential (\ref{p}) contains both the harmonic and the
inverse square interactions, for which the deformed Schr\"{o}dinger equation
has been solved exactly with the momentum representation (\ref{m}), see Refs.
\cite{boua1,chang}. However, the incorporation of the PHO interaction into
Schr\"{o}dinger equation in momentum space leads to a differential equation of
fourth order, for which the analytical solution is not known.

Given the importance of such potential, especially in molecular physics, we
investigate here its deformed Schr\"{o}dinger equation in coordinate space by
using the representation (\ref{brau}).

We proceed then by writing the Schr\"{o}dinger equation, for a particule with
reduced mass $\mu$ interacting with the PHO potential in position
representation as follows:%

\begin{equation}
\left(  \frac{\widehat{P}^{2}}{2\mu}+V(r)\right)  \psi(\overset{\rightarrow
}{r})=E\psi(\overset{\rightarrow}{r}). \label{5}%
\end{equation}

Using representation (\ref{brau}) in Eq. (\ref{5}), and neglecting terms of
order $\beta^{2}$, we obtain the following deformed Schr\"{o}dinger equation:
\begin{equation}
\left(  \frac{\widehat{p}^{2}}{2\mu}+V(r)+\frac{\beta}{\mu}\widehat{p}%
^{4}\right)  \psi(\overset{\rightarrow}{r})=E\psi(\overset{\rightarrow}{r}).
\label{6}%
\end{equation}

In the limit $\beta=0$, Eq. (\ref{6}) reduces to the ordinary Schr\"{o}dinger
equation. Its solution for this potential function can be found in the
standard textbooks of quantum mechanics, see, for instance, Ref.
\cite{landau}. For the bound state solutions, the energy eigenvalues and the
corresponding normalized eigenfunctions are \cite{R9,landau}:%

\begin{equation}
E_{n\ell}^{0}=-2D_{e}\left(  1-\frac{1}{\gamma}\left(  2n+1+\lambda\right)
\right)  ,\text{ \ \ \ }n=0,1,2,\ldots\label{E4}%
\end{equation}%
\begin{equation}
\psi_{n\ell m}^{0}(r,\theta,\varphi)=N_{n\ell}r^{-\frac{1}{2}+\lambda}%
\exp\left(  -\frac{\alpha}{2}r^{2}\right)  \ _{1}F_{1}\left(  -n,1+\lambda
;\alpha r^{2}\right)  Y_{\ell m}(\theta,\varphi),
\end{equation}
where $N_{n\ell}$ is a normalization constant given by%
\[
N_{n\ell}=\frac{\alpha^{\frac{1}{2}+\frac{\lambda}{2}}}{\Gamma(\lambda
+1)}\sqrt{\frac{2\Gamma(\lambda+n+1)}{n!}},
\]
$_{1}F_{1}$ is a confluent hypergeometric function, $Y_{\ell m}$ are the
spherical harmonics, $n$ and $\ell$ are, respectively, the radial
(vibrational) and orbiatal (rotational) quantum numbers. We have used the
notations
\[
\gamma=\ \sqrt{\frac{2\mu D_{e}r_{e}^{2}}{\hbar^{2}}}=\frac{4D_{e}}%
{\hbar\omega},\text{ }\lambda=\sqrt{\gamma^{2}+(\ell+\frac{1}{2})^{2}},\text{
\ }\alpha=\frac{\gamma}{r_{e}^{2}},\text{ \ \ }\omega=\frac{2}{r_{e}}%
\sqrt{\frac{2D_{e}}{\mu}},
\]
where $\omega$ represents the classical frequency for small harmonic vibrations.\ \ 

We return now to Eq. (\ref{6}). It is clearly seen that the effect of the
minimal length is described by the presence of the perturbation term
$\frac{\beta}{\mu}\widehat{p}^{4}$ in the ordinary Schr\"{o}dinger equation.
Thus, one can use the perturbation theory to compute the correction to the
energy levels in the first order in $\beta$. We can then write
\[
E_{n\ell}=E_{n\ell}^{0}+\Delta E_{n\ell},
\]
where $E_{n\ell}^{0}$ are the unperturbed levels corresponding to the
eigenfunctions $\psi_{n\ell m}^{0}(r)$, solutions to the ordinary
Schr\"{o}dinger equation, and $\Delta E_{n\ell}$ is the correction caused by
the minimal length. It is given by%
\[
\Delta E_{n\ell}=\frac{\beta}{\mu}\langle\psi_{n\ell m}^{0}\left\vert
p^{4}\right\vert \psi_{n^{\prime}\ell^{\prime}m^{\prime}}^{0}\rangle
\equiv\frac{\beta}{\mu}\langle n\ell m\left\vert p^{4}\right\vert n^{\prime
}\ell^{\prime}m^{\prime}\rangle.
\]
It has been shown in Ref. \cite{brau} that, for central interactions, $\Delta
E_{n\ell}$ can be expressed as follow:
\begin{equation}
\Delta E_{n\ell}=4\beta\mu\left[  \left(  E_{n\ell}^{0}\right)  ^{2}%
-2E_{n\ell}^{0}\langle n\ell m\left\vert V(r)\right\vert n\ell m\rangle
+\langle n\ell m\left\vert V^{2}(r)\right\vert n\ell m\rangle\right]  .
\label{8}%
\end{equation}

For the sake of later comparison with the harmonic oscillator, we write the
PHO potential (\ref{p}) in the general form%
\begin{equation}
V(r)=ar^{2}+\frac{b}{r^{2}}+c\text{ ,} \label{E2}%
\end{equation}
where the parameters $a$, $b$ and $c$ are related to the dissociation energy
$D_{e}$ and to the equilibrium internuclear distance $r_{e}$ by%
\[
a=D_{e}/r_{e}^{2},\text{ \ \ \ \ \ }b=D_{e}r_{e}^{2},\ \ \ \ \ \ c=-2D_{e}.
\]

By using formula (\ref{8}) with the PHO potential (\ref{E2}), we obtain
\begin{equation}%
\begin{array}
[c]{c}%
\Delta E_{n\ell}=4\mu\beta\left(  (E_{n\ell}^{0})^{2}+2ab+c^{2}\bigskip
-2cE_{n\ell}^{0}+\left(  2ac-2aE_{n\ell}^{0}\right)  \left\langle n\ell
m\right\vert r^{2}\left\vert n\ell m\right\rangle \right. \\
\left.  +a^{2}\left\langle n\ell m\right\vert r^{4}\left\vert n\ell
m\right\rangle +(2bc-2bE_{n\ell}^{0})\left\langle n\ell m\right\vert \frac
{1}{r^{2}}\left\vert n\ell m\right\rangle +b^{2}\left\langle n\ell
m\right\vert \frac{1}{r^{4}}\left\vert n\ell m\right\rangle \right)  .
\end{array}
\label{15}%
\end{equation}
To compute the minimal length correction, one has then to evaluate the matrix elements%

\[
\left\langle n\ell m\right\vert r^{q}\left\vert n\ell m\right\rangle
=N_{n,\ell}^{2}%
{\displaystyle\int\limits_{0}^{\infty}}
r^{q+1+2\lambda}\exp(-\alpha r^{2})\left[  _{1}F_{1}\left(  -n,1+\lambda
;\alpha r^{2}\right)  \right]  ^{2}dr,
\]
where $q=-4,-2,2,4$.

After evaluating the above integrals, we have obtained the following results:%

\begin{align*}
\left\langle n\ell m\right\vert r^{2}\left\vert n\ell m\right\rangle  &
=\frac{\lambda+2n+1}{\alpha},\\
\left\langle n\ell m\right\vert r^{4}\left\vert n\ell m\right\rangle  &
=\frac{(\lambda+1)(\lambda+2)+6n(\lambda+n+1)}{\alpha^{2}},\\
\left\langle n\ell m\right\vert \frac{1}{r^{2}}\left\vert n\ell
m\right\rangle  &  =\frac{\alpha}{\lambda},\\
\left\langle n\ell m\right\vert \frac{1}{r^{4}}\left\vert n\ell
m\right\rangle  &  =\alpha^{2}\frac{\lambda+2n+1}{\lambda(\lambda^{2}-1)}.
\end{align*}

By inserting these results in Eq. (\ref{15}), we arrive at the expression%
\begin{equation}%
\begin{array}
[c]{c}%
\Delta E_{n\ell}=4\mu\beta\left(  \left(  (E_{n\ell}^{0})^{2}+2ab+c^{2}%
\bigskip-2cE_{n\ell}^{0}\right)  +\sqrt{\dfrac{a\hbar^{2}}{2\mu}}%
(2c-2E_{n\ell}^{0})(\lambda+2n+1)+\right. \\
\left.  a\dfrac{\hbar^{2}}{2\mu}\left(  (\lambda+1)(\lambda+2)+6n(\lambda
+n+1)\right)  +\sqrt{\frac{2\mu}{\hbar^{2}}a}(2bc-2bE_{n\ell}^{0})\dfrac
{1}{\lambda}+\frac{2\mu}{\hbar^{2}}ab^{2}\dfrac{\lambda+2n+1}{\lambda
(\lambda^{2}-1)}\right)  .
\end{array}
\label{16}%
\end{equation}

Equation (\ref{16}) generalizes the formula given in Ref. \textrm{\cite{brau}}
for the harmonic oscillator potential, which is obtained when $b=c=0$. In this
case one has%

\[
\Delta E_{n\ell}^{ho}=4\beta\mu a\frac{\hbar^{2}}{2\mu}\left(  (\ell+\frac
{3}{2})(\ell+\frac{5}{2})+6n(\ell+n+\frac{3}{2})\right)  ,
\]
which is exactly the correction derived in Ref. \textrm{\cite{brau}.}

From Eqs.(\ref{E4}) and (\ref{16}), the complete energy spectrum of the PHO in
the presence of a minimal length can be written, in terms of the molecular
parameters $r_{e}$ and $D_{e}$, as folow:%

\begin{align}
E_{n\ell}  &  =E_{n\ell}^{0}+4\mu\beta\left(  (E_{n\ell}^{0})^{2}%
+4D_{e}E_{n\ell}^{0}+6D_{e}^{2}-(4D_{e}^{2}+2D_{e}E_{n\ell}^{0})\frac
{\lambda+2n+1}{\gamma}+\right. \nonumber\\
&  \left.  D_{e}^{2}\frac{\lambda^{2}+(6n+3)\lambda+6n(n+1)+2}{\gamma^{2}%
}-\gamma\frac{2D_{e}\left(  2D_{e}+E_{n\ell}^{0}\right)  }{\lambda}+D_{e}%
^{2}\gamma^{2}\frac{\lambda+2n+1}{\lambda(\lambda^{2}-1)}\right)  ,
\label{E12}%
\end{align}
where, $n=0,1,2,\ldots$, and $\lambda=\sqrt{\gamma^{2}+(\ell+\frac{1}{2})^{2}%
}.$

This formula is the main result of this work. It allows us to investigate the
effect of the GUP on the vibration-rotation energy levels of a given diatomic
molecule with the PHO interaction.

In fact, we may expand $E_{n,\ell}$ into powers of $1/\gamma$ since the
parameter $\gamma$ is so large ($\gamma\gg1$) for most molecules \cite{flug}.
We obtain after some calculations and simplifications the expression%

\begin{align}
E_{n\ell}  &  =4D_{e}\left(  (n+\frac{1}{2})\frac{1}{\gamma}+\frac{1}{4}%
(\ell+\frac{1}{2})^{2}\frac{1}{\gamma^{2}}\right) \nonumber\\
&  +4\mu\beta D_{e}^{2}\left\{  \left[  6n\left(  n+1\right)  +3\right]
\frac{1}{\gamma^{2}}+\left(  n+\frac{1}{2}\right)  \left[  4\left(  \ell
+\frac{1}{2}\right)  ^{2}+2\right]  \frac{1}{\gamma^{3}}\right\}  +O\left(
1/\gamma^{4}\right)  , \label{14}%
\end{align}
which can be approximated for a rotating diatomic molecules as follow:%

\begin{align}
E_{n\ell}  &  \approx4D_{e}(n+\frac{1}{2})\frac{1}{\gamma}+D_{e}(\ell+\frac
{1}{2})^{2}\frac{1}{\gamma^{2}}\nonumber\\
&  +(6\mu\beta D_{e}^{2})\frac{1}{\gamma^{2}}+24\mu\beta D_{e}^{2}(n+\frac
{1}{2})^{2}\frac{1}{\gamma^{2}}+16\mu\beta D_{e}^{2}\left(  n+\frac{1}%
{2}\right)  \left(  \ell+\frac{1}{2}\right)  ^{2}\frac{1}{\gamma^{3}}%
+O(\frac{1}{\gamma^{4}}). \label{17}%
\end{align}

This formula shows the effect of the deformed Heisenberg algebra on the energy
levels of a diatomic molecule interacting with the PHO potential. We can
observe that the minimal length correction carries new terms in the
ro-vibrational energy spectrum, which do not exist in the undeformed case,
such as the anharmonic vibrations and the coupling vibration-rotation. In
fact, the third term in Eq. (\ref{17}) can be viewed as a correction caused by
the minimal length to the dissociation energy of the molecule. The fourth term
is a correction that accounts for the anharmonicity of vibrations. Finally,
the last term is simply the effect of the minimal length on the coupling
vibration-rotation. However, it is easily seen, that the effect of the minimal
length on the harmonic vibrational energy levels is too small; the correction
of this energy is proportional to $1/\gamma^{3}$, and is neglected compared to
the first term in Eq. (\ref{17}). This result seems to be interesting as long
as the introduction of this elementary length brings all empirical terms of
diatomic molecules energy spectrum. Therefore, the PHO potential becomes a
more realistic model in this formalism of quantum mechanics with a minimal
length; its spectrum is now similar to those of the well known Kratzer and
Morse potentials \cite{flug}.

Furthermore, Formula (\ref{14}) can be viewed as an energy spectrum of a
three-parameter potential function, i.e., $D_{e},$ $r_{e},$ and $\beta$. In
this sense, the procedure of fitting used in the case of three-parameter
potentials, such as Morse function, can be followed for adjusting the
parameters of the "deformed PHO" with the spectroscopic data of diatomic
molecules. This viewpoint would be important because one can compute the
values of $\beta$ for any molecule, and estimate an order of magnitude of the
minimal length in molecular physics. For this, it would be interesting to
compute the energy spectrum in the general case ($\beta^{\prime}\neq2\beta$)
of the deformed Heisenberg algebra (\ref{4}). This is because, in the modified
Schr\"{o}dinger equation (\ref{6}), studied here, the perturbation term
$(\beta/\mu)\widehat{p}^{4}$can arise from many different sources, such as the
relativistic corrections, and hence it can describe not only the minimal
length. However, in the case ($\beta^{\prime}\neq2\beta$), the Schr\"{o}dinger
equation takes a complicated specific form, which clearly describe the minimal
length corrections. One can then, without any ambiguity, conclude about the
significance, and the order of the minimal length in diatomic molecules
physics.\textbf{ }

\section{Summary\ \ \ \ \ \ \ \ \ \ \ \ \ \ \ \ \ \ \ \ \ }

We studied the PHO potential in quantum mechanics with a GUP, which includes a
minimal length. By using a perturbative approach, the correction to the energy
spectrum of the PHO potential due to the existence of this fundamental length
is obtained analytically in coordinate space. The problem of
vibration-rotation of diatomic molecules is investigated in this modified
version of quantum mechanics by means of this potential model. The effect of
the minimal length on the ro-vibrational energy levels manifests by the
appearance of three kinds of corrections in the energy spectrum: the first
modifies the dissociation energy, the second concerns the anharmonic
vibrations, and the last correction affects the coupling vibration-rotation.
This result shows that the introduction of the minimal length improves this
molecular potential model since its energy spectrum includes now all the
necessary energy terms to describe the experimental vibration-rotation energy
levels of diatomic molecules. Therefore, It would be important to extend this
study to the general case of the deformed Heisenberg algebra (arbitrary values
of $\beta$ and $\beta^{\prime}$), and thereafter one can investigate
quantitatively the effect of the minimal length on the vibration-rotation of
diatomic molecules. This work is in progress and will be published else where.

\begin{acknowledgments}
This work was supported by the Algerian Ministry of Higher Education and
Scientific Research under the PNR Project No. 8/u18/4327 and the CNEPRU
Project No. D017201600026. We thank the referee for a pertinent remark that
led us to clarify the conclusion of this study.
\end{acknowledgments}

\end{document}